\documentclass[prc,twocolumn,superscriptaddress,superscriptfootnote,nofootinbib]{revtex4}
\usepackage{amssymb}

\usepackage{amsmath,amsfonts,amssymb,epsfig,color}

\begin{document}
\title{Symplectic symmetry and clustering in atomic nuclei}
\author{H. G. Ganev}
\affiliation{Joint Institute for Nuclear Research, Dubna, Russia}

\setcounter{MaxMatrixCols}{10}

\begin{abstract}
A new symplectic-based shell-model approach to clustering in atomic
nuclei is proposed by considering the simple system $^{20}$Ne. Its
relation to the collective excitations of this system is mentioned
as well. The construction of the Pauli allowed Hilbert space of the
cluster states with maximal permutational symmetry is given for the
$^{16}$O+$^{4}$He $\rightarrow$ $^{20}$Ne channel in the case of
one-component many-particle nuclear system. The equivalence of the
obtained cluster model space to that of the semi-microscopic
algebraic cluster model is demonstrated.
\\

\textbf{Keywords:} symplectic symmetry, clustering in nuclei,
algebraic symplectec-based shell-model approach
\end{abstract}
\maketitle

PACS number(s): {21.60.Fw; 21.60.Gx; 21.60.Ev}

\section{Collective and intrinsic excitations}

Probably the first authors who have pointed that the collective
excitations are related with the symplectic groups as
phenomenological dynamical groups are Goshen and Lipkin, which have
considered in detail the one- \cite{GL1} and two-dimensional
\cite{GL2} cases. After them, on the phenomenological level the
group $Sp(6,R)$ has been proposed as a dynamical group of collective
excitations in nuclear system by P. Raychev
\cite{Raychev72a,Raychev72b} and later considered in detail on
microscopic level by G. Rosensteel and D. Rowe \cite{RR77,RR80}.

From the other side, it was shown that the collective effects are
associated with operators that are scalar in $O(m)$ and the
collective Hamiltonian is obtained by projecting the many-particle
Hamiltonian on a definite $O(m)$ irrep associated with the $m$
relative Jacobi vectors in the configuration space
$\mathbb{R}^{3m}$, where $m=A-1$ and $A$ is equal to the total
number of nucleons in the system
\cite{Van74,Van76,Filippov81,Castanos82,GC}. By considering the
many-particle system as consisting of $A-1$ Jacobi quasiparticles
one readily avoids the problem of separation of the center-of-mass
motion from the very beginning. Asherova \emph{et al.}
\cite{Asherova75} have shown within the framework of the generalized
hyperspherical functions method (GHFM) that in its minimal
approximation, in which one restricts itself to a single $O(m)$
irreducible representation $(\omega_{1},\omega_{2}, \omega_{3})$ of
states of hyperspherical function within the lowest harmonic
oscillator shell, is equivalent to restricting to a single $SU(3)$
representation $(\omega_{1}-\omega_{2},\omega_{2}-\omega_{3})$. They
also showed, by means of an important $Sp(6,R) \otimes O(m)$
complementarity theorem of Moshinsky and Quesne \cite{MQ}, that the
diagonalization of the GHFM Hamiltonian in the minimal approximation
is equivalent to its diagonalization in the collective space of a
single $Sp(6,R)$ irreducible representation (irrep). In this way the
equivalence of the $O(m)$-based and symplectic-based theories of
nuclear collective motion was quickly realized by many authors, e.g.
\cite{Filippov81,Castanos82,GC,stretched,Van88}. Further, it was
shown in detail, e.g., by Filippov and collaborators
\cite{Filippov81} that the group $O(m)$ in the $Sp(6,R) \otimes
O(m)$ is related to the intrinsic motion of the relative Jacobi
quasiparticles with respect to the intrinsic principal axes frame of
the mass quadrupole tensor, on which the collective excitations are
built up. The group $O(m)$ is also of importance because, as we will
see further, it allows one to ensure the proper permutational
symmetry of the nuclear wave functions.

The above considerations can be reformulated more generally. We can
say that the set of basis states of the full dynamical symmetry
group $Sp(6m,R)$ of the whole many-particle nuclear system contains
different kinds of possible motions $-$ collective, intrinsic,
cluster, etc. However, often, one restricts himself to a certain
type of dominating excitation modes in the process under
consideration. Thus, by reducing the group $Sp(6m,R)$ one performs
the separation of the $3m$ nuclear many-particle variables $\{q\}$
into kinematical (internal) and dynamical (collective) ones, i.e.
$\{q\} = \{q_{D},q_{K}\}$. The choice of the reduction chain depends
on the concrete physical problem we want to consider. As we
mentioned, e.g., the group $Sp(6,R)$ plays an important role in the
treatment of the collective excitations in the one-component
many-particle nuclear system. The reduction $Sp(6m,R) \supset
Sp(6,R)\otimes O(m)$ thus turns out to be of a crucial importance in
the microscopic nuclear theory of collective motions. In this way
the considered reduction corresponds to the splitting of the
microscopic many-particle configuration space $\mathbb{R}^{3m}$,
spanned by the relative Jacobi vectors, into kinematical and
dynamical submanifolds, respectively. According to this, the
many-particle nuclear wave functions can be represented respectively
as consisting of collective and intrinsic components
\begin{equation}
\Psi = \sum_{\eta} \Theta_{\eta}(q_{D})\chi_{\eta}(q_{K}),
\label{WF}
\end{equation}
where $\chi_{\eta} \equiv | 0 \rangle =
|N_{0}(\lambda_{0},\mu_{0});KLM \rangle$ determines the bandhead
structure, and the collective function $\Theta_{\eta}$, in second
quantized form, can be written as a polynomial in the $Sp(6,R)$
raising operators.

As will see further, by considering another reduction chain of the
$Sp(6m,R)$ group one is able to isolate the cluster degrees of
freedom within the unified framework of the symplectic-based
shell-model approach to nuclear excitations. To be more specific, we
restrict ourselves to the case of two-cluster system only and
consider the simplest case of $^{20}$Ne system, which consists of
two structureless (closed-shell) $^{16}$O and $\alpha$ clusters.
Thus, for example, the intrinsic wave function of maximal space
symmetry which determines the intrinsic motion in $^{20}$Ne has the
$O(m)$ symmetry $(12,4,4)$. The collective excitations are then
generated by acting with the $Sp(6,R)$ raising generators on the
complementary $Sp(6,R)$ symplectic bandhead with $SU(3)$ symmetry
$(8,0)$. The $SU(3)$ basis states of the so obtained $Sp(6,R)$
irreducible representation for $^{20}Ne$ are given in Table
\ref{Ne20Sp6RIR}.

\begin{table}[h!]
\caption{The $Sp(6,R)$ representation $\langle \sigma \rangle =
\langle 12+\frac{19}{2}, 4+\frac{19}{2}, 4+\frac{19}{2} \rangle$ of
$^{20}Ne$.} \label{Ne20Sp6RIR}
\smallskip\centering\small\addtolength{\tabcolsep}{3.pt}
\begin{tabular}{lllll}
\hline &  & $\qquad\qquad\qquad\qquad \ \ \  \cdots $ &  &  \\
\hline
&  & $%
\begin{tabular}{l}
$\textcolor{red}{(12,0)},(10,1),2(8,2),(6,3),(7,1),(4,4),(6,0)$
\end{tabular}%
$ &  &  \\ \hline
&  & $%
\begin{tabular}{l}
$\qquad\qquad\qquad \textcolor{red}{(10,0)},(8,1),(6,2)$
\end{tabular}%
$ &  &  \\ \hline &  & $\qquad\qquad\qquad\qquad \ \ \
\textcolor{red}{(8,0)}$ &  &  \\ \hline
\end{tabular}%
\end{table}

\section{The $^{16}$O+$^{4}$He $\rightarrow$ $^{20}$Ne channel}

Using the simple $^{16}$O+$\alpha$ cluster system $^{20}$Ne, in
particular, we will prove the equivalence of the semi-microscopic
algebraic cluster model (SACM) \cite{SACMa,SACMb} and the
one-component symplectic-based scheme in the classification of the
cluster states in the many-particle nuclear Hilbert space.

\subsection{The semi-microscopic algebraic cluster model}

The semi-microscopic algebraic cluster model \cite{SACMa,SACMb} was
proposed as an approach to the cluster structure of light nuclei. In
the SACM, the relative motion of the clusters is described by the
vibron model \cite{VM}, whereas their internal structure is treated
in terms of the Elliott shell model, having a symmetry group
$U_{ST}(4) \otimes U_{C}(3)$, where $U_{C}(3)$ is the symmetry group
of the three-dimensional harmonic oscillator \cite{Elliott58} and
$U_{ST}(4)$ is the Wigner spin-isospin group \cite{Wigner37}. The
model space is constructed in a microscopic way by respecting the
Pauli principle, but the interactions are expressed in terms of
algebra generators. The states within the SACM for two-cluster
system are then classified by the following reduction chain
\cite{SACMa,SACMb}:
\begin{align}
&U_{ST,1}(4) \otimes U_{C1}(3) \otimes U_{ST,2}(4) \otimes U_{C2}(3)
\otimes U_{R}(4) \notag\\
\notag\\
&\supset U_{C_{1}}(3) \otimes U_{C_{2}}(3) \otimes U_{R}(3) \supset
U_{C}(3) \otimes U_{R}(3)  \notag\\
\notag\\
&\supset U(3) \supset SU(3) \supset SO(3) . \label{SACMchain}
\end{align}
The spin-isospin irreducible representations for $^{16}$O+$^{4}$He
$\rightarrow$ $^{20}$Ne channel are given by $\widetilde{f} =
[4444]$, $[1111]$ and $[5555]$, respectively. The $SU(3)$ intrinsic
structure for $^{16}$O and $\alpha$ clusters and combined system is
determined by the scalar, or equivalent to it, irreps of the groups:
$U_{C1}(3)$: $[4,4,4] \sim [0,0,0]$, $U_{C2}(3)$: $[0,0,0]$, and
$U_{C}(3)$: $[4,4,4] \sim [0,0,0]$. Then a complete set of quantum
numbers that characterize the cluster model states is defined by the
following chain:
\begin{align}
&U_{C_{1}}(3) \otimes U_{C_{2}}(3) \otimes U_{R}(3) \supset U_{C}(3)
\otimes U_{R}(3)  \notag\\
&\ [4,4,4] \quad [0,0,0] \ \ \ [n,0,0] \ \ \ [4,4,4] \ \ [n,0,0] \notag \\
\notag \\
&\supset \ \ U(3) \ \ \supset \ \ SU(3) \ \ \supset \ \ SO(3), \label{SACMchain} \\
&\ [4+n,4,4] \quad \ \ (n,0) \quad\quad\qquad L \notag
\end{align}
where for closed-shell nuclei the $SU_{ST,i}(4)$ groups were
dropped. Thus, the cluster model space is spanned by the $SU(3)$
irreps $(n_{0}+n,0)$, where $n = 0, 1, 2, \ldots$. The Pauli
principle requires $n_{0} = 8$.

\subsection{The one-component symplectic-based approach}

At this point we want to point out once again that the cluster
excitations, together with the collective and intrinsic motions, are
naturally contained in the full dynamical group $Sp(6m,R)$ of the
many-nucleon nuclear system. In the case of two-cluster system
($A=A_{1}+A_{2}$), the well-known anzatz \cite{Wildermuth77}
\begin{equation}
\Psi = \mathcal{A}\{\phi_{1}(A_{1}-1)\phi_{2}(A_{2}-1)f(q_{0})\}
\label{WFRGM}
\end{equation}
can be related to the symplectic scheme by considering the following
reduction chain:
\begin{align}
&Sp(6(A-1),R) \notag\\
\notag\\
&\supset Sp(6A_{1},R) \otimes Sp(6(A_{2}-1),R) \notag\\
\notag\\
&\supset Sp(6,R)_{0} \otimes Sp(6(A_{1}-1),R) \otimes
Sp(6(A_{2}-1),R), \label{op2}
\end{align}
where the groups $Sp(6(A_{1}-1),R)$ and $Sp(6(A_{2}-1),R)$ describe
the intrinsic state of the fist and second cluster, respectively.
One of the $(A-1)$ Jacobi vectors, denote it by $q_{0}$ (from the
set $A_{1}$), will describe the relative motion of the two clusters,
whereas the rest $(A-2)$ Jacobi vectors will be related to the
intrinsic states of the clusters. Thus, the group $Sp(6,R)_{0}$ in
(\ref{op2}) will describe the "cluster excitations", related to the
relative distance vector $q_{0}$. The group $Sp(6(A_{1}-1),R)$,
describing the first cluster states, can be further reduced to
$Sp(6,R)_{A_{1}-1} \otimes O(A_{1}-1)$, i.e. $Sp(6(A_{1}-1),R)
\supset Sp(6,R)_{A_{1}-1} \otimes O(A_{1}-1)$, where as usual the
first group in the direct product describes the collective
excitations, whereas the second one $-$ the intrinsic state of the
cluster, on which the collective excitations are built up. Due to
the mutually complementarity relationship \cite{MQ}, the
$Sp(6,R)_{A_{1}-1}$ irrep $\langle \sigma \rangle$ is completely
determined by the irrep $\omega$ of $O(A_{1}-1)$ and vice versa.
Similar considerations are valid for the second cluster with
$(A_{2}-1)$ Jacobi quasiparticles. The wave function (\ref{WFRGM})
in this case can be easily obtained from Eq.(\ref{WF}) by
identifying $\chi_{\eta}(q_{K}) = \phi_{1}(A_{1}-1)
\phi_{2}(A_{2}-1)$ and $\Theta_{\eta}(q_{D})=f(q_{0})$.

To relate the present classification scheme to that of SACM we
consider further the reduction of the subgroups in Eq.(\ref{op2}),
i.e. the complete reduction chain:
\begin{align}
&Sp(6(A-1),R) \notag\\
\notag\\
&\supset Sp(6,R)_{0} \otimes Sp(6(A_{1}-1),R) \otimes Sp(6(A_{2}-1),R) \notag\\
\notag\\
&\supset Sp(6,R)_{0} \otimes Sp(6,R)_{A_{1}-1} \otimes O(A_{1}-1) \notag\\
\notag\\
&\qquad\qquad\quad \ \otimes Sp(6,R)_{A_{2}-1} \otimes O(A_{2}-1) \notag\\
\notag\\
&\supset U(3)_{0} \otimes U_{A_{1}-1}(3) \otimes U_{A_{2}-1}(3)
\otimes S_{A_{1}} \otimes S_{A_{2}} \notag\\
&\quad [n,0,0] \quad \ \ [4,4,4] \quad\quad \ [0,0,0] \quad\quad
f_{1} \quad \ \ f_{2} \notag\\
\notag\\
&\supset \ \ U(3) \ \ \supset \ \ SU(3) \ \ \supset \ \ SO(3) . \label{sp6mR-2cc}\\
&\ [4+n,4,4] \quad\quad (n,0) \qquad\quad \ \ L \notag
\end{align}
The permutational symmetry of the clusters is given by the
reductions: $O(15) \downarrow S_{16}$: $(4,4,4) \rightarrow
\{4,4,4,4\}$ and $O(3) \downarrow S_{4}$: $(0) \rightarrow \{4\}$.
Then the permutational symmetry of the combined systems will be $f =
\{4,4,4,4,4\}$. Because of the full antisymmetry of the total wave
function, the spin-isospin content of the combined system is given
by conjugate Young scheme $\widetilde{f} = [5,5,5,5]$. The cluster
model space for the $^{20}$Ne two-cluster system is then spanned by
the even and odd $Sp(6,R)_{0}$ irreps respectively, i.e. by the sets
of $SU(3)$ irreps: $(8,0), (10,0), (12,0), \ldots$ and $(9,0),
(11,0), (13,0), \ldots$, which constitute a representation of the
double covering metaplectic group $Mp(6,R)$. Alternatively, the two
$Sp(6,R)_{0}$ irreps could be unified into a single irrep of the
semidirect product group $[HW(3)_{0}]Sp(6,R)_{0}$. Note that because
the $Sp(6,R)_{0}$ irreps are built up by means of a single Jacobi
vector $q_{0}$, corresponding to the intercluster distance, only
one-rowed irreps of the subgroup $SU(3)_{0} \subset Sp(6,R)_{0}$ are
allowed of the type $(n_{0}+2n,0)$. Thus, the positive-parity
cluster state space in the $^{16}$O+$^{4}$He $\rightarrow$ $^{20}$Ne
channel within the one-component symplectic-based scheme will
coincide with the $Sp(6,R)$ irreducible collective space that is
spanned by the \emph{fully symmetric $SU(3)$ irreducible
representations only}, given in Table \ref{Ne20Sp6RIR} in red.
Correspondingly, the full cluster model space with maximal
permutational symmetry of the SACM given by the $SU(3)$ irreps
$(n_{0}+n,0)$ with $n = 0, 1, 2, \ldots$ is obtained by considering
both the even and odd irreducible collective spaces of
$Sp(6,R)_{0}$, in which the states of the three-dimensional harmonic
oscillator of even and odd number of oscillator quanta fall. Thus,
the cluster model spaces of the SACM and the one-component
symplectic-based approach to the cluster states are identical. It is
then clear that, based on the equivalence of the microscopic model
spaces, the usage of the same (algebraic) Hamiltonian in both the
SACM and one-component symplectic-based schemes to the clustering in
atomic nuclei will produce identical spectra. We note also that the
role of the Wigner supermultiplet group $SU_{ST,i}(4)$ ($i=1,2$),
which is important in the construction of the Pauli allowed model
space, is played in the symplectic-based scheme by the orthogonal
group $O(A_{i}-1)$ through its reduction $O(A_{i}-1) \supset
S_{A_{i}}$.

The present work also concerns the important question of the mutual
interrelations between the shell model, symplectic $Sp(6,R)$ model
and cluster model states. A microscopic cluster model has been
introduced in 1958 by K. Wildermuth and T. Kanellopoulos
\cite{Wildermuth58} who have showed its relation to the shell model.
Soon this relation has been reformulated in terms of the $SU(3)$
symmetry by B. F. Bayman and A. Bohr \cite{Bayman58}. In this way it
has been recognized very early the important role played by the
$SU(3)$, and complementary to it $SU(4)$, symmetry in clustering
phenomena in atomic nuclei. The connection of the shell model and
the cluster model states has later been given in algebraic terms by
K. T. Hecht \cite{SM-cluster} using $SU(3)$ and $SU(4)$ coupling and
recoupling techniques. The relation of the $Sp(6,R)$ and
$\alpha$-cluster model states has been done in
Refs.\cite{Sp6R-cluster1,Sp6R-cluster2,Sp6R-cluster3}. Recall that
the symplectic classification of the nuclear states organizes the
Hilbert space of the nucleus vertically into vertical cones or
slices (called irreducible collective spaces), in contrast to the
conventional shell model in which the Hilbert space is organized
horizontally into different shells or layers. In
Ref.\cite{Sp6R-cluster1,Sp6R-cluster2} it has been demonstrated that
the $\alpha$-cluster and $Sp(6,R)$ states are essentially
complementary with decreasing overlap with the increase of the
oscillator quanta excitations $2n \hbar\omega$ and may both be
needed for a meaningful microscopic description of light nuclei. The
cluster states obtained in the present paper actually coincide with
the so called stretched $SU(3)$ states \cite{stretched} of the type
$(\lambda_{0}+2k,\mu_{0})$ with $k=0, 1, 2, \ldots$ of the $Sp(2,R)
\subset Sp(6,R)$ submodel \cite{Sp2R} representing the core
collective excitations along the $z$-direction only. The connection
of the $\alpha$-cluster and $Sp(2,R)$ states has been investigated
in Refs.\cite{Sp2R-cluster1,Sp2R-cluster2} and \cite{Sp2R-cluster3}
for the case of $^{8}$Be and $^{12}$C, respectively. Recently, the
no-core symplectic model (NCSpM) with the $Sp(6,R)$ symmetry has
been used to study the many-body dynamics that gives rise to the
ground state rotational band together with phenomena tied to
alpha-clustering substructures in the low-lying states in  $^{12}$C
\cite{Dreyfuss13,Launey16}. The intersection of the shell,
collective and cluster models of the atomic nuclei has also been
given in a similar algebraic perspective in
Refs.\cite{Cseh15,Cseh21} through the consideration of the common
dynamical algebraic structure $SU_{x}(3) \otimes SU_{y}(3)$ of all
the three fundamental models of nuclear structure, where $x$ stands
for the bandhead (valence shell), whereas $y$ refers to the
major-shell excitations.

\section{Conclusions}

In the present paper, a new symplectic-based shell model approach to
clustering in atomic nuclei is proposed. The cluster degrees of
freedom are isolated by an appropriate separation of the full set of
relative Jacobi many-particle variables into dynamical (collective)
and kinematical (intrinsic) ones by reducing the full dynamical
symmetry group $Sp(6m,R)$ of the whole one-component (no distinction
is made between the proton and neutron degrees of freedom)
many-particle nuclear system. According to this, the nuclear wave
functions are represented as having cluster (collective) and
intrinsic components. The kinematical part allows to ensure all the
integrals of motion of the considered nuclear system, including the
proper permutational symmetry. The symplectic symmetry thus provides
the nuclear cluster systems with fully microscopic shell-model wave
functions that respect the Pauli principle.

For simplicity, the proposed algebraic approach is illustrated for
the case of two-cluster nuclear systems by considering the simple
system $^{20}$Ne. The construction of Pauli allowed Hilbert space of
the cluster states with maximal permutational symmetry is worked out
for the $^{16}$O+$^{4}$He $\rightarrow$ $^{20}$Ne channel in the
case of one-component many-particle nuclear system. The equivalence
of the obtained cluster model space to that of the semi-microscopic
algebraic cluster model is demonstrated. Further all the
symplectic-based computational machinery can be used in practical
applications. In contrast to the semi-microscopic algebraic cluster
model, the symplectic $Sp(6,R)$ symmetry allows to build up the
required quadrupole collectivity observed in some nuclei without the
use of an effective charge. Thus, the present approach can further
be tested in obtaining the excitation spectrum, including the
microscopic structure of the cluster states and the transition
probabilities, not only in $^{20}$Ne, but also in other light nuclei
for which the clustering is supposed to play an important role.

The relation of the present symplectic-based shell model approach to
the collective excitations in $^{20}$Ne is mentioned as well. We
note also that the cluster motion is a relative motion of the one
group of nucleons in phase with respect to the another group, i.e.
it is also of collective nature. In this regard, the cluster motion
appears as a specific kind of collective excitations. Indeed, as we
have demonstrated in the present work, the cluster model state space
is a restricted part of the corresponding irreducible collective
space of the two-cluster system as a whole.

Finally, the algebraic approach presented here could be generalized
for more clusters or/and to the case of the two-component
proton-neutron cluster nuclear systems. The extension to more
clusters, however, requires the proper consideration of the point
group symmetries. The role of the different point symmetries in
clustering have recently been studied very successfully by R.
Bijker, F. Iachello and their collaborators
\cite{Bijker14,D3h-12C,Bijker19} within the framework of the
algebraic cluster model (ACM) with $U(\nu+1)$ dynamical group of
$\nu=3(k-1)$ degrees of freedom for nuclei composed of $k$
$\alpha$-particles \cite{Bijker20}. The point symmetries can be
taken into account in the present symplectic-based shell model
approach to clustering by considering the reduction of the
orthogonal group $O(A-1)$ and its subgroup $S_{A} \subset O(A-1)$.
For the two-cluster systems ($A=A_{1}+A_{2}$), subject of the
present work, one needs to consider the following two reduction
chains: $O(A-1) \supset O(A_{1}-1) \oplus O(A_{2}-1) \oplus O(1)$
and $S_{A} \supset S_{A_{1}} \oplus S_{A_{2}} \oplus S_{2}$. In this
case, the geometry of the intercluster motion described by a single
Jacobi vector $q_{0}$ is determined by the reduction $O(1) \supset
S_{2}$ of the trivial orthogonal group $O(1)$, consisting of two
discrete elements $\{\pm1\}$. The cluster states of the two-cluster
system are therefore classified by the scalar $S_{2}$ irrep $\{2\}$
obtained according to the $O(1)\downarrow S_{2}$ branching rule.
Taking the isomorphism $S_{2} \sim Z_{2}$, one sees that this
classification corresponds to a dumbbell configuration with $Z_{2}$
symmetry. Similarly, for the case of three-cluster system
($A=A_{1}+A_{2}+A_{3}$), the point symmetry can be taken into
account by considering the following reduction chains: $O(A-1)
\supset O(A_{1}-1) \oplus O(A_{2}-1) \oplus O(A_{3}-1) \oplus O(2)$
and $S_{A} \supset S_{A_{1}} \oplus S_{A_{2}} \oplus S_{A_{3}}
\oplus S_{3}$. Then the geometry of the three-cluster system, which
intercluster motion is described by two Jacobi vectors $q_{01}$ and
$q_{02}$, can be classified by the reduction $O(2) \supset S_{3}$
associated with these two Jacobi vectors. The branching rules for
$O(2)\downarrow S_{3}$ are as follows: $(0)\downarrow \{3\}$,
$(0)^{\star}\downarrow \{111\}$, $(3a+p)\downarrow \{3\}+\{111\}$
for $p=0$, and $\{21\}$ for $p=1,2$. In this way, the three-cluster
system states can be classified according to the different
irreducible representations of $S_{3} \sim D_{3h}$, corresponding to
an equilateral triangle configuration with $D_{3h}$ symmetry.

\end{document}